\documentclass[twocolumn]{aastex701}
\received{ }
\revised{ }
\accepted{ }
\submitjournal{ApJ}
\begin{document}

\title{A Possible Stellar CME Event Associated with a Long-duration Optical Flare on the\\ RS CVn-type Star UX~Arietis}

\affiliation{International Centre of Supernovae (ICESUN), Yunnan Key Laboratory of Supernova Research, Yunnan Observatories, Chinese Academy of Sciences, Kunming 650216, China}
\affiliation{Yunnan Observatories, Chinese Academy of Sciences, Kunming 650216, China}

\author[orcid=0000-0002-3534-1740]{Dongtao Cao}
\affiliation{International Centre of Supernovae (ICESUN), Yunnan Key Laboratory of Supernova Research, Yunnan Observatories, Chinese Academy of Sciences, Kunming 650216, China}
\affiliation{Key Laboratory for the Structure and Evolution of Celestial Objects, Chinese Academy of Sciences, Kunming 650216, China}
\email[show]{dtcao@ynao.ac.cn}  

\author[orcid=0009-0004-3025-3444]{Shenghong Gu}
\affiliation{Yunnan Observatories, Chinese Academy of Sciences, Kunming 650216, China}
\affiliation{Key Laboratory for the Structure and Evolution of Celestial Objects, Chinese Academy of Sciences, Kunming 650216, China}
\affiliation{School of Astronomy and Space Science, University of Chinese Academy of Sciences, Beijing 101408, China}
\email[show]{shenghonggu@ynao.ac.cn}

\correspondingauthor{Dongtao Cao, Shenghong Gu} 

\begin{abstract}
Coronal mass ejections (CMEs) are large-scale eruptions of magnetized plasma from stellar coronae, and they are always accompanied by flares. In this study, we present high-resolution spectroscopic observations of a long-duration optical flare on the RS CVn-type star UX Arietis (UX Ari). Around the flare peak, we have detected a far-blueshifted emission component in the H$\alpha$ line profile---a signature of an erupting prominence---indicating a possible stellar CME. After correcting for projection effects, the eruption velocity reaches $\sim -432$~km~s$^{-1}$ when assuming radial propagation, significantly exceeding the escape velocity of the primary component ($\approx 304$~km~s$^{-1}$). This strongly supports the detection of a stellar CME. The mass of this CME is at least $2.7 \times 10^{20}$~g, and its kinetic energy is $2.5 \times 10^{35}$~erg. The mass value is consistent with the solar flare--CME scaling relation, while its kinetic energy lies below the solar trend. The flare itself released an energy of $4.2 \times 10^{36}$~erg in the H$\alpha$, corresponding to a white-light bolometric energy of roughly $2.8 \times 10^{38}$~erg. Throughout the flaring event, evolving asymmetries in the H$\alpha$ profile reveal complex plasma dynamics.
\end{abstract}



\section{Introduction}\label{sec1}
Stellar flares, which are explosive phenomena driven by the sudden releases of magnetically stored energy in stellar coronae, have been widely observed and documented in cool stars across a broad range of wavelengths from radio to X-ray bands. A point of particular interest is superflares with energies between $10^{33}$ and $10^{38}$~erg \citep{Schaefer2000}, exceeding the strongest solar flare on the record \citep[$\sim10^{32}$~erg, e.g.,][]{Emslie2012}. Such powerful events are thought to be closely linked to the triggers of more energetic stellar coronal mass ejections (CMEs). 

As a well-studied reference, CMEs on the Sun are large-scale eruptions of magnetized plasma, typically originating from the destabilization of magnetic flux ropes in the corona \citep[e.g.,][]{Chen2011} and are often associated with magnetic reconnection \citep[e.g.,][]{Schmieder2015}. This reconnection process also powers solar flares, establishing a close physical connection between flares and CMEs: flares release energy via magnetic reconnection in closed magnetic loops, while CMEs open up the magnetic field and expel plasma and magnetic flux into the interplanetary medium \citep[e.g.,][]{Priest2002}. In many cases, a prominence/filament---a cool, dense plasma structure suspended in the corona---erupts as part of a CME event, providing a clear observational signature in chromospheric lines such as the H$\alpha$ \citep[e.g.,][]{Gopalswamy2003}. Understanding this flare--prominence/filament--CME relationship on the Sun provides a crucial framework for identifying analogous events on other active stars.

For stars, stellar CMEs can erode planetary atmospheres and alter their composition through enhanced ionizing radiation and energetic particle fluxes, thereby critically affecting planetary habitability \citep{Airapetian2016, Cherenkov2017, Hazra2022}. Moreover, frequent CMEs may contribute significantly to stellar mass and angular momentum losses, playing an important role in stellar evolution \citep{Aarnio2012, Osten2015}.

%------------------------------------------------------------------------------------------%
\begin{deluxetable}{lccc}
\tablenum{1}
\tablecaption{Basic Stellar Parameters of UX~Ari\label{tab1}}
\tablewidth{\linewidth}
%\tabletypesize{\normalsize}
\setlength{\tabcolsep}{15pt} 
\tablehead{
\colhead{Parameter} & \colhead{Primary} & \colhead{Secondary} & \colhead{Ref.}
}
\startdata
\hline
Spectral Type & K0~IV & G5~V & \citet{Carlos1971} \\
$T_{\mathrm{eff}}$ (K) & 4560~$\pm$~100 & 5670~$\pm$~100 & \citet{Hummel2017} \\
$\log g$ (cgs) & 3.06 & 4.09 & \citet{Hummel2017} \\
Mass (M$_{\odot}$) & 1.30~$\pm$~0.06 & 1.14~$\pm$~0.06 & \citet{Hummel2017} \\
Radius (R$_{\odot}$) & 5.6~$\pm$~0.1 & 1.6~$\pm$~0.2 & \citet{Hummel2017} \\
$vsini$ (km~s$^{-1}$) & 39 & 7.5 & \citet{Cao2017} \\
$P_{\mathrm{orb}}$ (days) & \multicolumn{2}{c}{6.437888~$\pm$~0.000007} & \citet{Hummel2017} \\
$P_{\mathrm{rot}}$ (days) & \multicolumn{2}{c}{6.437888~$\pm$~0.000007} & \citet{Hummel2017} \\
Distance (pc) & \multicolumn{2}{c}{50.5} & \citet{Gaia2020} \\
Inclination ($^{\circ}$) & \multicolumn{2}{c}{59.5} & \citet{Duemmler2001} \\
\hline
\enddata
\end{deluxetable}
%------------------------------------------------------------------------------------------%

The RS CVn-type binary system comprises at least one cooler component that exhibits particularly intense magnetic activity in various forms; for instance, it is capable to produce flares that are several orders of magnitude more energetic than solar flares \citep[e.g.,][]{Inoue2023, Cao2024, Cao2025}. Given the well-established correlation between highly energetic flares and CMEs observed on the Sun, therefore, it can be hypothesized that RS CVn-type stars may experience frequent and intense CME occurrences. \citet{Inoue2023} reported the detection of a high-velocity prominence eruption that resulted in a CME event associated with a superflare on the RS CVn-type star V1355~Orionis. Furthermore, \citet{Cao2024} identified a potential flare-associated CME candidate on the RS CVn-type star II~Peg through a prominent redshifted H$\alpha$ component of 429 km~s$^{-1}$, a speed exceeding the escape velocity of II~Peg. Additional CME candidates on II~Peg, detected via flare-associated asymmetries in chromospheric line profiles, were reported based on long-term high-resolution spectroscopic monitoring \citep{Cao2025ApJ}. And more recently, \citet{Cao2025} documented a failed filament eruption at the onset of a long-duration optical flare on the RS CVn-type star UX~Ari.

UX~Ari (=~HD~21242~=~BD+28$^{\circ}$532) is a well-studied, non-eclipsing spectroscopic binary system in a nearly circular orbit ($P \approx 6.44$~days), consisting of a K0~IV primary component and a G5~V secondary component \citep{Carlos1971, Hummel2017}. Table~\ref{tab1} summarizes the basic parameters of UX~Ari. UX~Ari is a highly active star, showing significant starspot activity \citep[e.g.][]{Raveendran1995, Aarum2003a, Rosario2008, Xiang2026}, and strong chromospheric emission in the H${\alpha}$, $\mbox{Ca~{\sc ii}}$ H \& K, and $\mbox{Ca~{\sc ii}}$ IRT lines \citep[e.g.][]{Carlos1971, Nations1986, Huenemoerder1989, Raveendran1995, Montes1996, Montes2000, Gu2002, Aarum2003b}. It is widely acknowledged that the chromospheric emission is mainly ascribed to the K0~IV primary component. In addition, UX~Ari is also a high-rate flaring star, exhibiting flares many times in the wide range of wavelength \citep{Simon1980, Elias1995, Montes1996, Franciosini2001, Gu2002, Richards2003, Aarum2003b, Cao2017, Kurihara2024, Cao2025, Urabe2025, Inoue2026}.

In this study, we present the results of the time-resolved high-resolution spectroscopic observations of UX~Ari. In Section~\ref{sec2}, we provide details of our observation, data reduction, and spectral subtraction of chromospheric activity lines. Analysis and results of flaring spectra are presented in Section~\ref{sec3}. The discussion is given in Section \ref{sec4}. Finally, we conclude the study by summarizing our new findings in Section \ref{sec5}.

%------------------------------------------------------------------------------------------%
\section{Spectroscopic observation, data reduction, and spectral subtraction}\label{sec2}
The high-resolution spectroscopic observation of UX~Ari were carried out with the fiber-fed High-Resolution Spectrograph (HRS) mounted on the 2.16~m telescope at the Xinglong station of the National Astronomical Observatories, Chinese Academy of Sciences \citep{Fan2016}. The observation took place from 2015 October~27 to November 2. The HRS produces spectra with a resolving power R~=~$\lambda$/$\Delta\lambda$~$\simeq$~48000 in a wavelength range of 3900--10000~\AA, using a $4096 \times 4096$ pixel CCD detector. The detailed observing information can be found in Table~\ref{tab2}.

The data reduction was performed with the IRAF\footnote{IRAF is distributed by the National Optical Astronomy Observatories, which is operated by the Association of Universities for Research in Astronomy (AURA), Inc., under cooperative agreement with the National Science Foundation.} package, following the prescription described in \citet{Cao2024}. With an interactive procedure in the IRAF package, the heavy telluric lines appeared in the chromospheric lines were removed by using a telluric template derived from the spectrum of the brighter and rapidly rotating early-type star HR~8858 (Spectral type: B5~V, $vsini$ = 316~km~s$^{-1}$). 

\begin{figure}
\centering
\includegraphics[width=8.5cm,height=10.0cm]{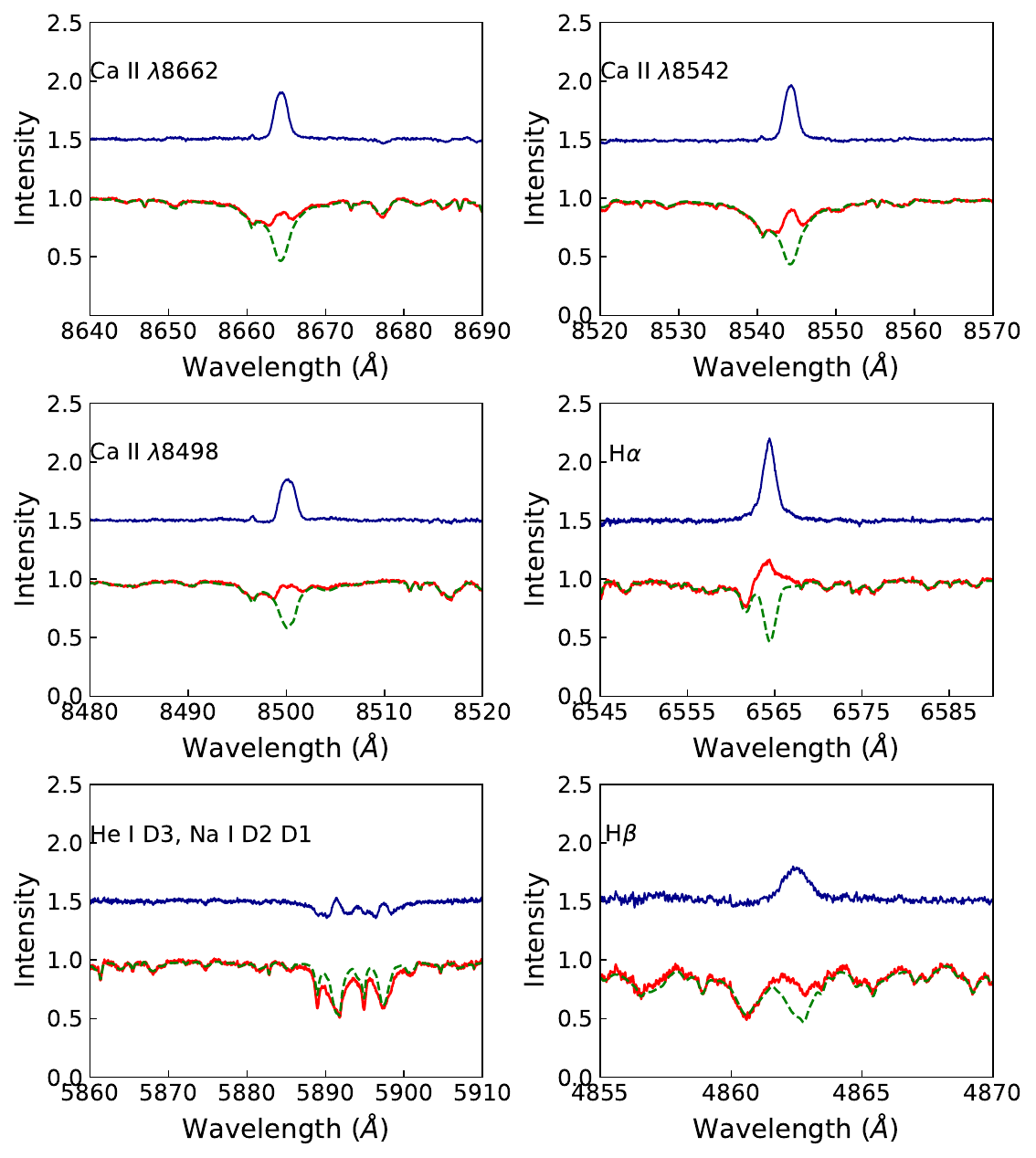}
\caption{Examples of observed, synthesized, and subtracted spectra for the $\mbox{Ca~{\sc ii}}$~$\lambda$8662, $\mbox{Ca~{\sc ii}}$~$\lambda$8542, and $\mbox{Ca~{\sc ii}}$~$\lambda$8498, H${\alpha}$, $\mbox{Na~{\sc i}}$ D$_{1}$, D$_{2}$ doublet, $\mbox{He~{\sc i}}$ D$_{3}$, and H${\beta}$ spectral regions, from one spectrum obtained at phase 0.350 on 2015 November~1. In each panel, the lower solid red line indicates the observed spectrum, the dashed green line represents the synthesized spectrum constructed from spectra of two template stars, and the upper darkblue line shows the resulting subtraction spectrum, shifted for better visibility. The label identifying each chromospheric line is also marked.}
\label{Fig1}
\end{figure}

\begin{figure*}
\centering
\includegraphics[width=14.25cm,height=16.75cm]{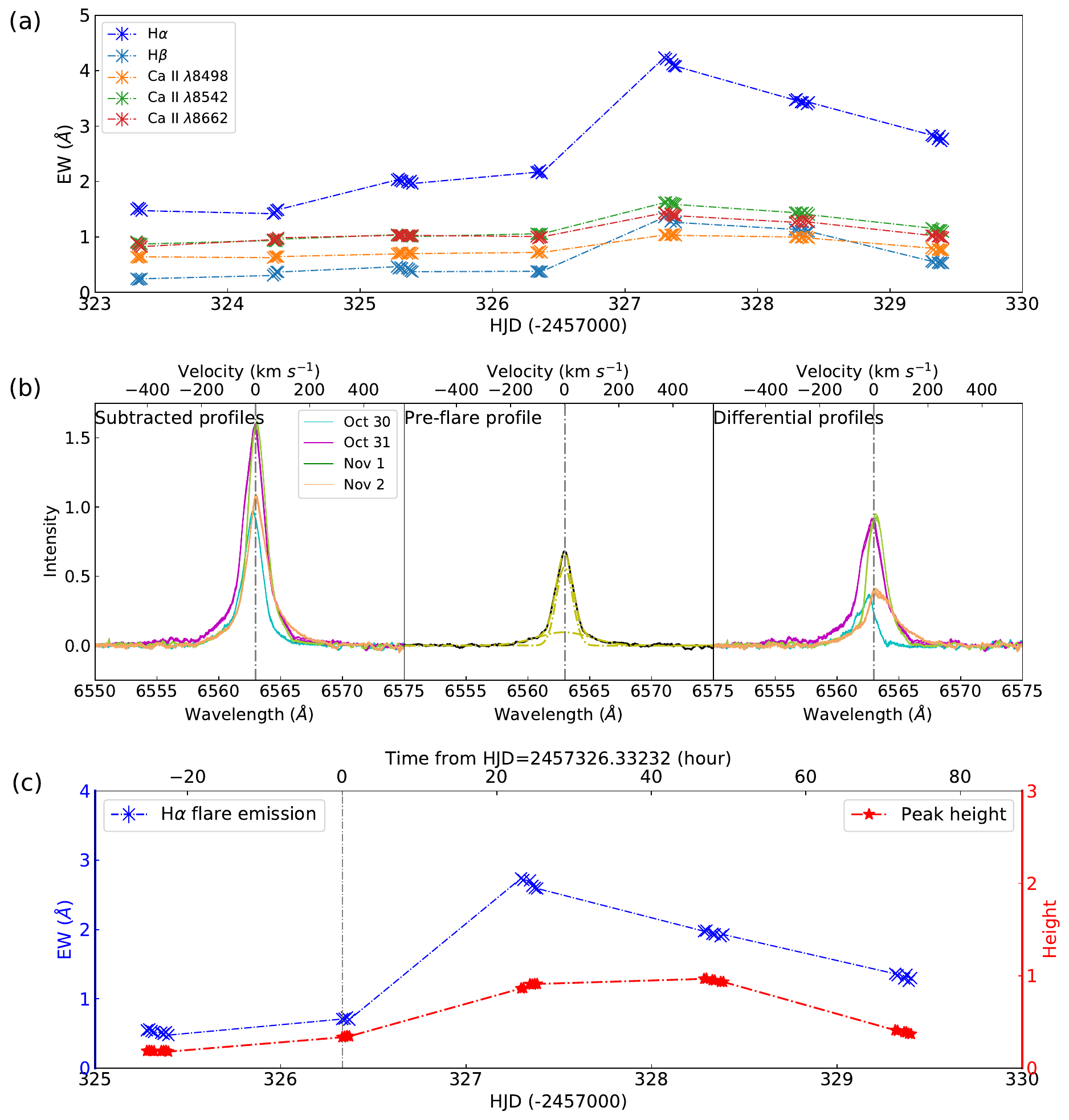}
\caption{Temporal variations of the indicators' EWs and H${\alpha}$ line profile. Top panel: The EWs of the STARMOD-subtracted $\mbox{Ca~{\sc ii}}$ $\lambda$8662, $\mbox{Ca~{\sc ii}}$ $\lambda$8542, $\mbox{Ca~{\sc ii}}$ $\lambda$8498, H${\alpha}$, and H${\beta}$ line profiles plotted against the HJD. Middle panel: The STARMOD-subtracted H${\alpha}$ profiles, the pre-flare H${\alpha}$ profile, and the differential H${\alpha}$ profiles. All spectra are corrected to the rest velocity frame of the K0 IV primary star of UX Ari. Bottom panel: EWs (left axis) and peak heights (right axis) of the H$\alpha$ line after removing the intrinsic chromospheric contribution, plotted against the HJD and flare time (upper axis). The zero point of flare time ($t = 0$) corresponds to HJD2457326.33232, marked by the vertical dash-dotted line.}
\label{Fig2}
\end{figure*}

UX~Ari is a spectroscopic binary system in which the spectral lines of the two components exhibit Doppler shifts and blending as the orbital phase varies. To isolate the chromospheric activity signature from the observed line profiles and thereby enable a robust analysis of activity-related line profile features, we therefore apply the spectral subtraction technique by the STARMOD program \citep{Barden1985, Montes1997, Montes2000}. The STARMOD synthesizes a spectrum by rotationally broadening, radial velocity (RV)-shifting, and adaptively weighting spectra of two appropriate template stars. These template stars are selected for their inactivity while sharing the same spectral type and luminosity class as the components of the target binary system. The synthesized spectrum serves as an approximation of the non-active state of the binary, and therefore the subtraction between the observed and synthesized spectra yields a representation of the activity contribution. Here, we use spectra of HD~3351 (K0~IV) and HR 3309 (G5 V) as templates for the primary and secondary components of UX~Ari. The $vsini$ values of 39~km~s$^{-1}$ for the primary component and 7.5~km~s$^{-1}$ for the secondary component are used, and the intensity weight ratios of two components are 0.77/0.23 for the $\mbox{Ca~{\sc ii}}$ $\lambda$8662 spectral region, 0.765/0.235 for the $\lambda$8542 spectral region, 0.76/0.24 for the $\mbox{Ca~{\sc ii}}$ $\lambda$8498 spectral region, 0.74/0.26 for the H${\alpha}$ spectral region, 0.69/0.31 for the $\mbox{He~{\sc i}}$ D$_{3}$ spectral region, and 0.66/0.34 for the H${\beta}$ spectral region. As an example, we show the chromospheric activity lines of one spectrum and illustrate the above described processing in Figure~\ref{Fig1}. The activity contribution is quantified by measuring the equivalent widths (EWs) of the subtracted chromospheric line profiles, and the resulting EWs are plotted against the HJD in Figure~\ref{Fig2}(a). 
%------------------------------------------------------------------------------------------%
%------------------------------------------------------------------------------------------%
%------------------------------------------------------------------------------------------%
\begin{figure*}
\centering
\includegraphics[width=18cm,height=22cm]{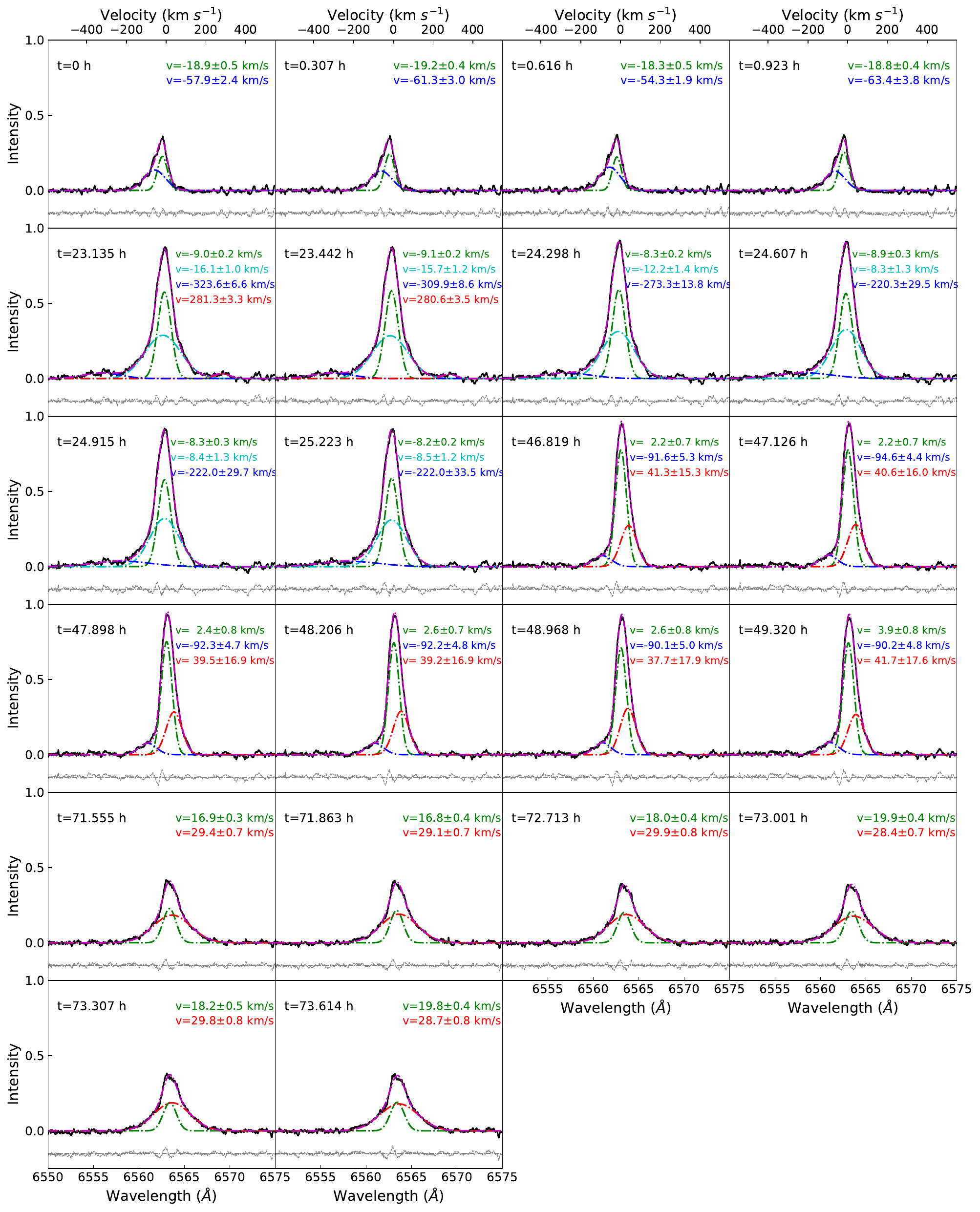}
\caption{Differential H${\alpha}$ profiles and their Gaussian fits. In each panel, the differential profile is shown as a black solid line and the best-fit model as a magenta dash-dotted line. Individual Gaussian components are plotted as colored dashed lines. Residuals between the differential and modeled spectra are displayed beneath each profile.}
\label{Fig3}
\end{figure*}
%------------------------------------------------------------------------------------------%

\section{Analysis and Results}\label{sec3}
\subsection{Optical flare(s)}
Clear emission feature in the $\mbox{He~{\sc i}}$ D$_{3}$ line---a well-established flare indicator due to its high excitation potential---was observed on 2015 October 30, 31, and November 1. During the same period, other chromospheric activity lines, most notably H$\alpha$, also exhibited enhanced emission. These signatures indicate that the spectra obtained on those nights are flaring, resulting from either a single long-duration optical flare or a series of independent flare events. By 2015 November 2, the $\mbox{He~{\sc i}}$ D$_{3}$ line no longer showed obvious emission, whereas H$\alpha$ and other chromospheric lines continued to display strong emission. This pattern likely corresponds either to the most decay phase of the preceding long-duration flare, or to a separate, weaker optical flare event.

The STARMOD-subtracted spectra during the flare(s) contain both intrinsic chromospheric activity emission and flare emission. To isolate the pure flare radiation and analyze spectral asymmetries relative to the flaring activity, we first correct the subtracted spectra to the rest frame of the K0~IV primary star. We then construct a pre-flare reference profile by averaging the subtracted H$\alpha$ profiles from 2015 October 27 (HJD2457323) and 28 (HJD2457324), when the system showed lower chromospheric emission. The observation from 2015 October 29 was excluded because the elevated emission likely corresponds to a precursor phase of the flare. Subtracting this pre-flare profile from each flaring spectrum yields the differential spectra shown in Figure~\ref{Fig2}b, from which the pure flare H$\alpha$ radiation is extracted (Figure~\ref{Fig2}c).

Table~\ref{tab3} presents the EWs, surface fluxes, and luminosities derived from the differential H$\alpha$ spectra. The calculation of surface fluxes and luminosities follows the methodology of \citet{Cao2025}.

%-----------------------------------------------------%
\begin{deluxetable}{lccc}
\tablenum{2}
\tablecaption{EWs, Surface Fluxes, and Flare Luminosities of the Differential H$\alpha$ Spectra\label{tab3}}
\tablewidth{0pt}
\tablehead{
\colhead{Phase}&\colhead{EW}&\colhead{Fs ($\times~10^{6}$)}&\colhead{L ($\times~10^{30}$)}\\
\nocolhead{}&\colhead{({\AA})}&\colhead{(erg~cm$^{-2}$~s$^{-1}$)}&\colhead{(erg~s$^{-1}$)} 
}
\startdata
\multicolumn4c{2015 Oct 30}\\
0.030&0.708$\pm$0.005&2.734&5.212\\
0.032&0.714$\pm$0.001&2.757&5.256\\
0.034&0.731$\pm$0.004&2.823&5.381\\
0.036&0.699$\pm$0.013&2.699&5.145\\
\hline
\multicolumn4c{2015 Oct 31}\\
0.180&2.739$\pm$0.012&10.576&20.161\\
0.182&2.716$\pm$0.015&10.487&19.991\\
0.188&2.710$\pm$0.001&10.464&19.947\\
0.190&2.631$\pm$0.018&10.159&19.366\\
0.192&2.602$\pm$0.024&10.047&19.152\\
0.194&2.593$\pm$0.010&10.012&19.085\\
\hline
\multicolumn4c{2015 Nov 1}\\
0.333&1.976$\pm$0.001&7.630&14.545\\
0.335&1.985$\pm$0.001&7.664&14.610\\
0.340&1.939$\pm$0.002&7.487&14.272\\
0.342&1.932$\pm$0.001&7.460&14.221\\
0.347&1.904$\pm$0.001&7.352&14.015\\
0.350&1.928$\pm$0.004&7.444&14.190\\
\hline
\multicolumn4c{2015 Nov 2}\\
0.493&1.361$\pm$0.007&5.255&10.017\\
0.495&1.341$\pm$0.004&5.178&9.871\\
0.501&1.294$\pm$0.013&4.996&9.524\\
0.503&1.346$\pm$0.001&5.197&9.907\\
0.505&1.259$\pm$0.015&4.861&9.266\\
0.507&1.299$\pm$0.003&5.016&9.562\\
\enddata
\end{deluxetable}
%-----------------------------------------------------%
%------------------------------------------------------------------------------------------%

\subsection{H${\alpha}$ line profiles during the flare(s)}
As shown in Figure~\ref{Fig2}(b), it is notable that the pre‑flare H$\alpha$ profile is clearly symmetric. Therefore, the asymmetric features seen in the differential H$\alpha$ spectra are intrinsic to the flare rather than arising from the subtraction process. We have modeled these differential profiles with multi‑component Gaussian fits, as displayed in Figure~\ref{Fig3}; the corresponding fitting results are also presented in the same figure.

During the interval \(t = 0\)--\(0.9\)~hr, the differential H$\alpha$ profiles are well fitted with two Gaussian components: a broad, blueshifted emission component with an average bulk velocity of approximately $-59$ km~s$^{-1}$, and a narrower, slightly blueshifted emission component averaging roughly at $-19$ km~s$^{-1}$.

\begin{figure}
\centering
\includegraphics[width=8.5cm,height=6.cm]{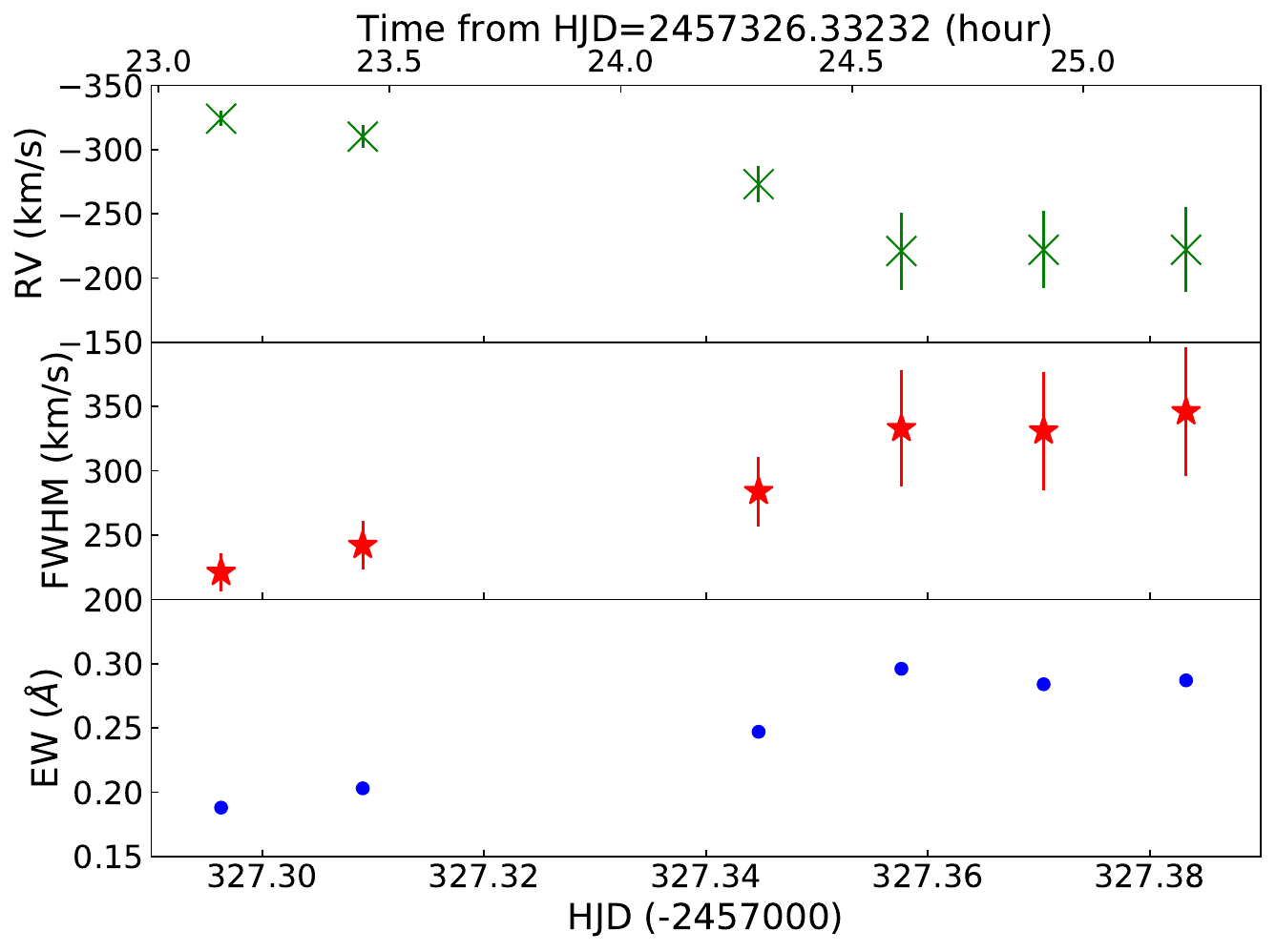}
\caption{Temporal evolution of the RV (top panel), FWHM (middle panel), and EW (bottom panel) of the far-blueshifted emission component observed during the interval from \(t = 23.1\) to \(25.2\)~hr.}
\label{Fig5b}
\end{figure}

Between \(t = 23.1\)--\(25.2\)~hr, the differential H$\alpha$ profiles exhibit more complex structure. The first two profiles are successfully reproduced with four Gaussian components. In addition to a nearly central broad component and a narrow component, we identify a highly blueshifted emission feature with a bulk velocity exceeding $-300$ km~s$^{-1}$ and a weaker, highly redshifted component with a bulk velocity of about $280$ km~s$^{-1}$. For the subsequent four spectra, the redshifted component weakens and disappears, allowing us to adopt a three‑Gaussian model. Notably, during this period the blueshifted component gradually decelerates to about $222$ km~s$^{-1}$ and then stabilizes near this velocity, while its full width at half‑maximum (FWHM) increases. The temporal evolution of the velocity, FWHM, and EW of this far‑blueshifted emission component is shown in Figure~\ref{Fig5b}.

%------------------------------------------------------------------------------------------%
%------------------------------------------------------------------------------------------%
In the period \(t = 46.8\)--\(49.3\)~hr, the differential H$\alpha$ profiles require three Gaussian components for accurate fits: a blueshifted emission component averaging at about $-92$ km~s$^{-1}$, a nearly central component around $3$ km~s$^{-1}$, and a redshifted component at roughly 40 km~s$^{-1}$.

Finally, during \(t = 71.6\)--\(73.6\)~hr, the differential H$\alpha$ profiles are best fitted with two Gaussian components: a narrow emission component averaging at approximately 18 km~s$^{-1}$ and a notably broad emission component with an average bulk velocity of about 29 km~s$^{-1}$.
%------------------------------------------------------------------------------------------%
\begin{figure}
\centering
\includegraphics[width=8.5cm,height=5.5cm]{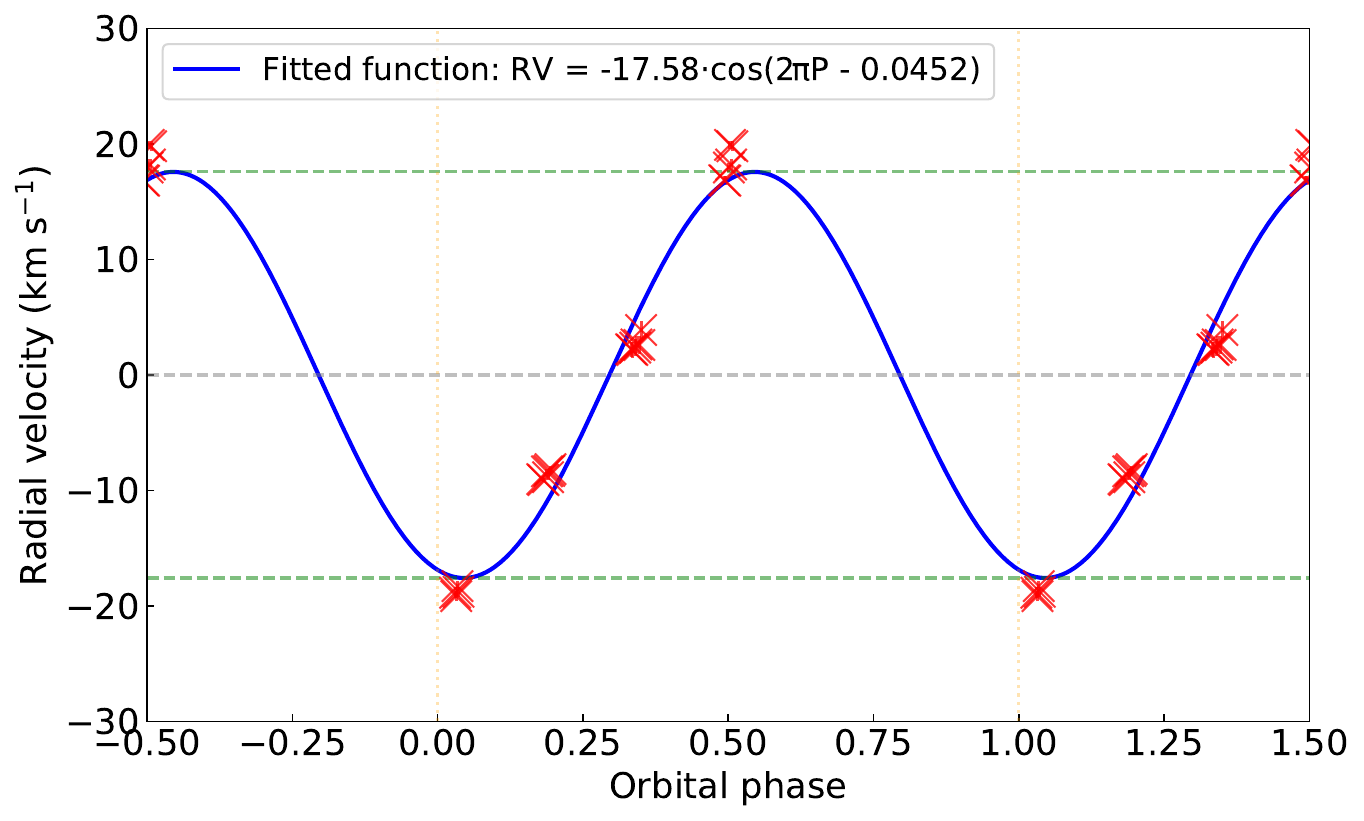}
\caption{Bulk RVs of the central emission component in the differential H$\alpha$ profiles (red crosses) and the best-fit sinusoidal curve (blue line). The horizontal green dashed lines indicate the amplitude of the sinusoidal fit. This sinusoidal variation reflects corotation of the flaring region with the stellar surface.}
\label{Fig4}
\end{figure}
%------------------------------------------------------------------------------------------%

\section{Discussion}\label{sec4}
\subsection{A long-duration flare on UX~Ari}
Unlike many active M‑type dwarf stars (e.g., EV~Lac and YZ~CMi) that frequently produce clusters of short white‑light flares in Transiting Exoplanet Survey Satellite (TESS) photometry, UX~Ari does not exhibit such frequent flaring activity. Instead, it has shown evidences of long‑duration flare activity. For instance, visual inspection of Monitor of All-sky X-ray Image (MAXI) data reveals an X‑ray flare lasting from MJD59078.3 to MJD59079.8 ($\sim$1.5 days). More recently, an optical flare reported by \citet{Cao2025} persisted for at least 150 hours. In our case, as shown in Figure~\ref{Fig2}c, the H$\alpha$ emission variation itself follows a long-duration, flare-like profile. Furthermore, modeling of the differential line profiles indicates that the asymmetric components correspond to flare‑related dynamic phenomena, while the nearly central narrow emission component likely originates from the flaring region itself. The velocity shift of this central narrow component is naturally explained by the corotation of that flaring region with the stellar surface—a configuration strongly indicative of a long-lived event. To test this interpretation, we have fitted the RV variation of this component with a sinusoidal function. The observed motion closely matches the expected trend, confirming the corotation scenario. These lines of evidence support our idea that the flare signatures described in Section~3.1 all originate from a single long-duration event that persisted for at least 73.6 hours (the limit of our observing coverage).
%------------------------------------------------------------------------------------------%
\begin{figure}
\centering
\includegraphics[width=9cm,height=10cm]{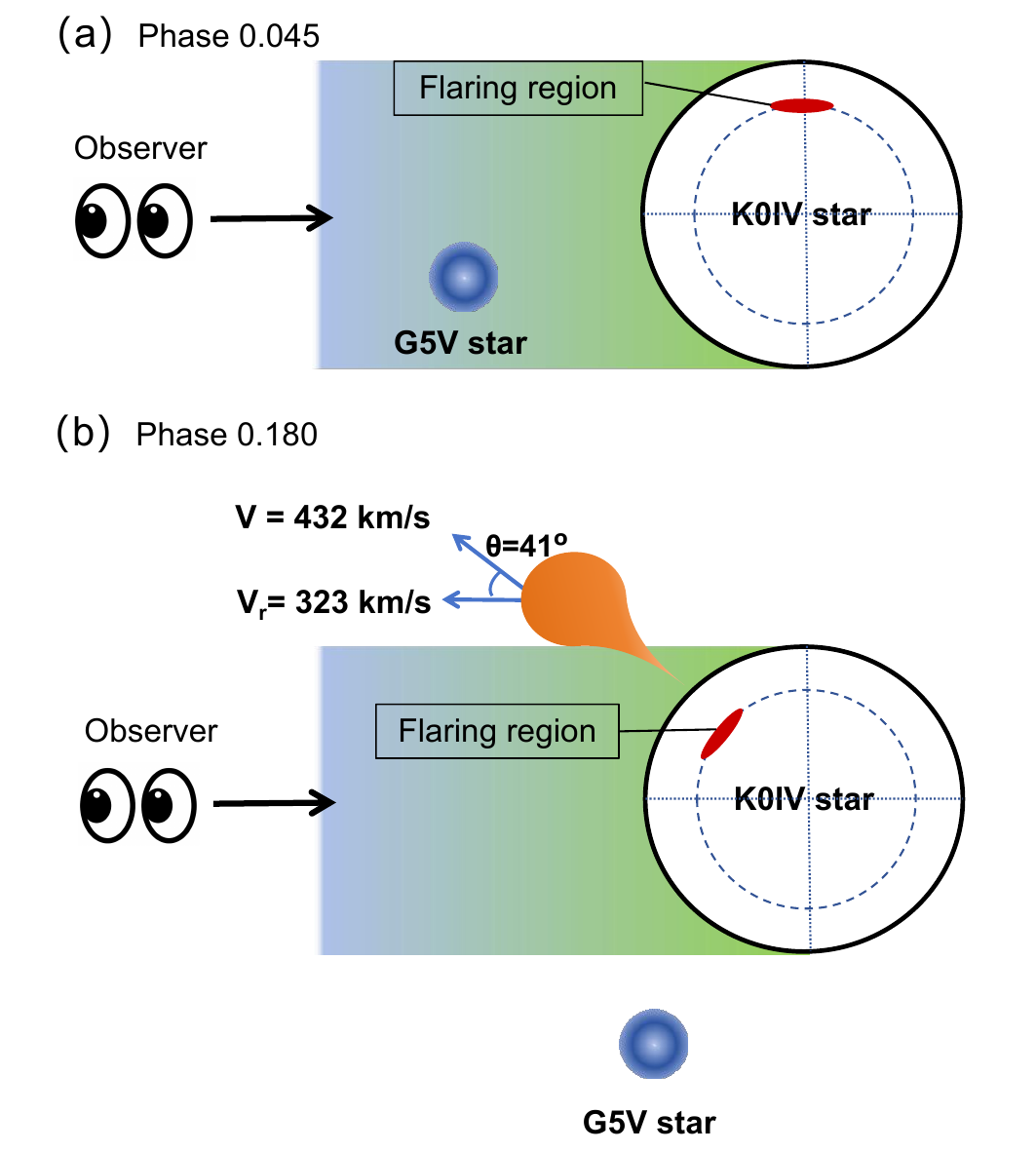}
\caption{Schematic diagram that expresses the interpretation of the possible flaring location. Panel (a): Flaring region on the K0\,IV primary star at orbital phase 0.045, where its radial velocity toward the observer is at maximum. Panel (b): Flaring region and the associated CME geometry at observing phase 0.18, showing the line-of-sight projection effect. The dashed circles represent the latitude of 63$^{\circ}$.}
\label{Fig5}
\end{figure}
%------------------------------------------------------------------------------------------%

The long-duration flare should originate from the active K0~IV primary star of UX~Ari, given the star’s known high activity level and the clear association of the STARMOD-subtracted H$\alpha$ emission with this component during the flare. From the sinusoidal fit to the RV of central emission component, we can estimate the approximate location of the flaring region. Using the fitted amplitude of approximately 17.58~km~s$^{-1}$ together with the star’s projected rotational velocity ($vsini$ = 39~km~s$^{-1}$), the latitude of the flaring region is inferred to be roughly $63^\circ$. It should be noted, however, that the H\(\alpha\) profiles used for the fitting originate from the chromosphere, whereas the projected rotational velocity is derived from the photospheric lines. Given that the chromosphere may rotate at a slightly different rate than the photosphere, the inferred latitude should be regarded as an approximate estimate rather than a precise measurement. Furthermore, the fitted sinusoidal curve reaches its maximum radial velocity toward the observer at the orbital phase~0.045 (see the schematic in Figure~\ref{Fig5}a). According to the system’s ephemeris, this corresponds to the flaring region being located at an orbital phase of approximately 0.705 on the K0~IV primary star. Notably, this inferred location is consistent with the starspot distribution derived from the Doppler imaging by \citet{Xiang2026}, who found that spot regions on UX Ari are predominantly concentrated at a latitude of $60^\circ$ and around orbital phases 0.25 and 0.75. %The close agreement in latitude and the proximity to one of the active longitudes strongly suggest that the flaring region was specially associated with the starspot region on the stellar surface.

Having established the flare's origin and location, we quantify its energy release. To estimate the total energy released by this long-duration optical flare, we adopt the method described in \citet{Cao2025}. We first compute the mean nightly H$\alpha$ flare luminosity, integrate it over 24 hr for each night, and then sum the contributions across all nights. This yields a total flare energy of $\sim 4.2 \times 10^{36}$~erg in the H$\alpha$ line. Using the H$\alpha$-to-bolometric energy scaling relation from Equation~(2) of \citet{Namekata2024}, we have converted this to an estimated bolometric (white-light) flare energy of $\sim 2.8 \times 10^{38}$~erg. This value lies within the energy range typical for stellar superflares, which are commonly defined as events releasing energies $\gtrsim 10^{33}$~erg.

\subsection{Interpretations of the H$\alpha$ Line Profile Asymmetries}
Asymmetries in chromospheric line profiles during flares provide valuable diagnostics for identifying events such as stellar prominence eruptions and potential CMEs \citep[e.g.][]{Vida2019, Koller2021, Namekata2022, Wu2022, Lu2022, Inoue2023, Cao2024, Cao2025ApJ, Cao2026}. It should be noted, however, that stellar prominence eruptions or CMEs are not the only sources of such line-profile asymmetries; other flare-related plasma motions can also produce similar observational signatures. %Insights from solar flare studies indicate that a blueshifted asymmetry in the line profile often originates from chromospheric evaporation, typically with upward velocities of several tens of km~s$^{-1}$ \citep[e.g.,][]{Tei2018}. Conversely, a redshifted emission component can arise from several distinct processes: chromospheric condensation, with downward velocities also on the order of several tens of km~s$^{-1}$ \citep{Ichimoto1984}; the downward draining of coronal rain along post-flare loops, which generally descends at speeds between 30 and 200~km~s$^{-1}$, averaging around 60--70~km~s$^{-1}$ \citep{Antolin2012, Lacatus2017}; or the infall of material in a failed prominence eruption \citep{Koller2021, Wu2022}.

During the initial flare phase \(t = 0\)--\(0.9\) hr, the broad blueshifted H$\alpha$ component at \(-59\)~km~s$^{-1}$ is most readily explained by chromospheric evaporation. In this process, intense flare heating drives rapid upward plasma flows from the lower atmosphere into the corona, producing the observed blueshifted emission. The measured velocity falls within the typical range of several tens of~km~s$^{-1}$ seen in solar chromospheric evaporation \citep[e.g.,][]{Tei2018}, and its appearance during the early energy-release stage further supports this interpretation. While chromospheric evaporation provides a natural explanation, an alternative scenario—a prominence eruption—could also produce a blueshifted signature. In typical eruptive events, the velocity of an erupting prominence tends to increase rapidly over time. In contrast, the blueshifted component in our spectra remains nearly constant during this interval. The absence of significant acceleration argues against an erupting prominence in the usual sense. However, we cannot entirely rule out the possibility that the flare initiated a slow, pre-eruptive rise of a prominence that had not yet developed into a full eruption. Such a scenario could also produce a steady, low-velocity blueshift.

The observations at \(t = 23.1\)--\(25.2\)~hr correspond to the peak phase of the flare. Aside from a narrow, nearly central emission component, a significantly broadened component is present---most likely attributable to strong Stark broadening under high electron densities of this evolutionary stage. Moreover, a high‑velocity blueshifted emission component emerges with a bulk velocity of \(-323\)~km~s\(^{-1}\) in the initial spectra of this interval. This high-velocity blueshifted component is likely caused by an erupting prominence. Notably, this velocity slightly exceeds the estimated surface escape velocity of the K0\,IV primary star, $v_{esp}~=~630(\frac{M_{\star}}{M_{\sun}})^{1/2}(\frac{R_{\star}}{R_{\sun}})^{-1/2}$~km~s$^{-1}$~$\approx$~304 km~s$^{-1}$, indicating that the prominence eruption may develop into a stellar CME. It should be noted that the measured RV is a lower limit due to projection effects. Based on the inferred flaring location (orbital phase \(\sim 0.705\)) and the observing phase (\(0.180\)), we estimate a projection angle of \(\sim 41^\circ\) between the line of sight and the eruption direction in longitude. Incorporating the flare latitude (\(\sim 63^\circ\)) and the stellar inclination (\(\sim 59.5^\circ\)) yields a deprojected velocity of \(\sim -432\)~km~s\(^{-1}\) (see Figure~\ref{Fig5}b). This corrected velocity substantially exceeds the escape speed, providing stronger support for the CME interpretation. We caution that this deprojected true velocity assumes a radial eruption from the flaring region, whereas solar CMEs often exhibit non-radial propagation, particularly in their early phases \citep[e.g.][]{Cremades2004}. Throughout this interval, this high-velocity blueshifted component gradually decelerated while its FWHM increased (see Figure~\ref{Fig5b}). This behavior closely resembles the evolution of solar CMEs, which typically decelerate after initial acceleration and expand as they propagate outward, leading to spectral line broadening. The observed velocity and FWHM evolution probably provides further evidences that this blueshifted component traces a stellar CME propagating outward from the star.

During the interval from \(t = 46.8\) to \(49.3\) hours, the flare entered a gradual decay phase. Although the EWs decreased from flare peak values, the line profile heights did not show a comparable decline (see Figure~\ref{Fig2}c), suggesting that a secondary flare might occur and the energy release was still ongoing. Therefore, the observed blueshifted component (\(\sim -92\) km~s\(^{-1}\)) is likely attributable to chromospheric evaporation, while the redshifted component (\(\sim 40\) km~s\(^{-1}\)) probably originates from chromospheric condensation. Solar flare studies indicate that condensation downflows are typically on the order of tens of km~s\(^{-1}\) \citep[e.g.,][]{Ichimoto1984}, consistent with our measured velocity, supporting a solar--stellar analogy. In addition, an alternative interpretation is coronal rain---plasma cools, condenses, and falls back along post-flare loops under gravity. Coronal rain becomes more pronounced during the decay phase, with velocities typically ranging from \(30\) to \(200\) km~s\(^{-1}\) (average \(\sim 60\)--\(70\) km~s\(^{-1}\); \citealt{Antolin2012, Lacatus2017}), making it a plausible alternative explanation for the redshifted component.

During the late decay phase (\(t = 71.6\)--\(73.6\)~hr), the flaring region had rotated near the stellar limb. The H\(\alpha\) profile is best fitted by two Gaussian components. The narrow component arises from the flaring region itself, now viewed at the limb, and traces cooling, quiescent plasma in post-flare loops. The broad, slightly redshifted component is also likely attributable to coronal rain. The resulting redshifted signal is accompanied by a significant line broadening, which can be explained by velocity dispersion during downfall, turbulent motions, and enhanced projection of tangential motions at the limb geometry.

Throughout this long-duration flare event, the evolving H\(\alpha\) profiles reveal a complex sequence of dynamical phenomena, including chromospheric evaporation, a stellar CME, condensation downflows, and coronal rain, highlighting the diverse plasma processes associated with energetic stellar flare and demonstrating the unique diagnostic power of time-resolved high-resolution spectroscopy.

\subsection{Properties of the stellar CME and their implications}
We estimate a lower limit on the mass of the detected stellar CME following the methods of \citet{Houdebine1990} and \citet{Koller2021} using the expression:
\begin{equation}
M_{\mathrm{CME}} \geqslant \frac{4\pi d^{2} f_{\mathrm{line}} m_{\mathrm{H}}}{A_{ji} h \nu_{ji} P_{\mathrm{esc}}} \frac{N_{\mathrm{tot}}}{N_{j}},
\end{equation}
where \(d\) is the distance to the star, \(f_{\mathrm{line}}\) is the line flux of the observed asymmetric feature, and \(m_{\mathrm{H}}\) is the mass of a hydrogen atom. The ratio \(N_{\mathrm{tot}}/N_{j}\) gives the total hydrogen number density relative to the population in the excited level \(j\); \(h\), \(\nu_{ji}\), and \(A_{ji}\) denote Planck’s constant, the transition frequency, and the Einstein coefficient for spontaneous decay from level \(j\) to level \(i\), respectively; and \(P_{\mathrm{esc}}\) is the escape probability. Specific parameter values are adopted from \citet{Koller2021}. We have computed the luminosity of the asymmetric features, replacing the term \(4\pi d^{2} f_{\mathrm{line}}\), which yields a CME mass of about \(2.7\times10^{20}\)~g. The kinetic energy of the CME is given by \(E_k = \frac{1}{2} M_{\mathrm{CME}} v^2\), where \(v\) is the eruption velocity. This mass value together with the bulk line‑of‑sight velocities of the asymmetric features gives a corresponding kinetic energy of approximately \(1.4\times10^{35}\) erg. When the deprojected true eruption velocity is adopted, the kinetic energy increases to about \(2.5\times10^{35}\) erg.
%This mass together with the bulk line‑of‑sight velocities of the asymmetric features gives a corresponding kinetic energy of approximately \(1.4\times10^{35}\) erg. When the deprojected true eruption velocity is adopted, the kinetic energy increases to about \(2.5\times10^{35}\) erg.

\begin{figure}
\centering
\includegraphics[width=8.5cm,height=5.75cm]{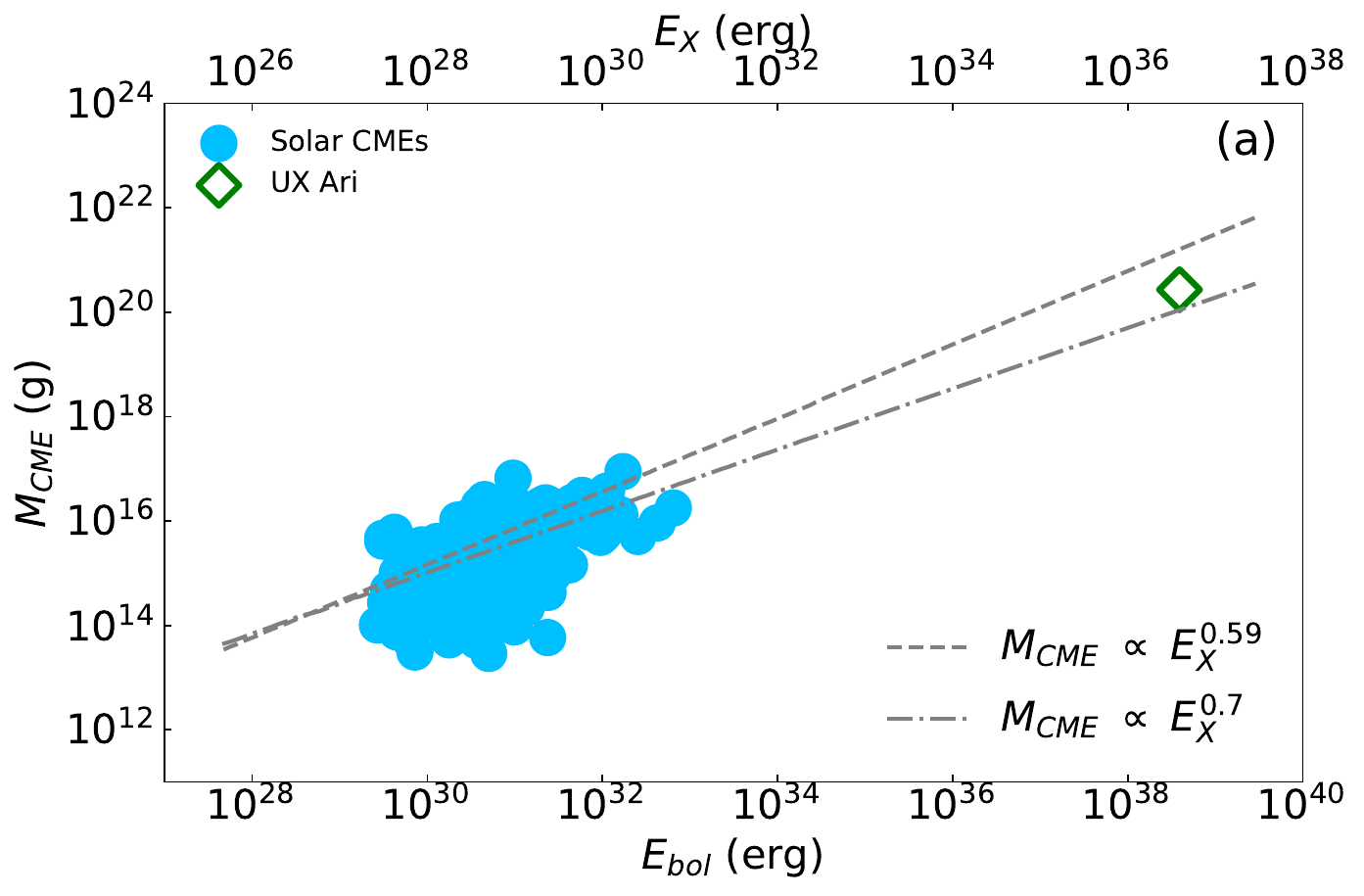}
\includegraphics[width=8.5cm,height=5.75cm]{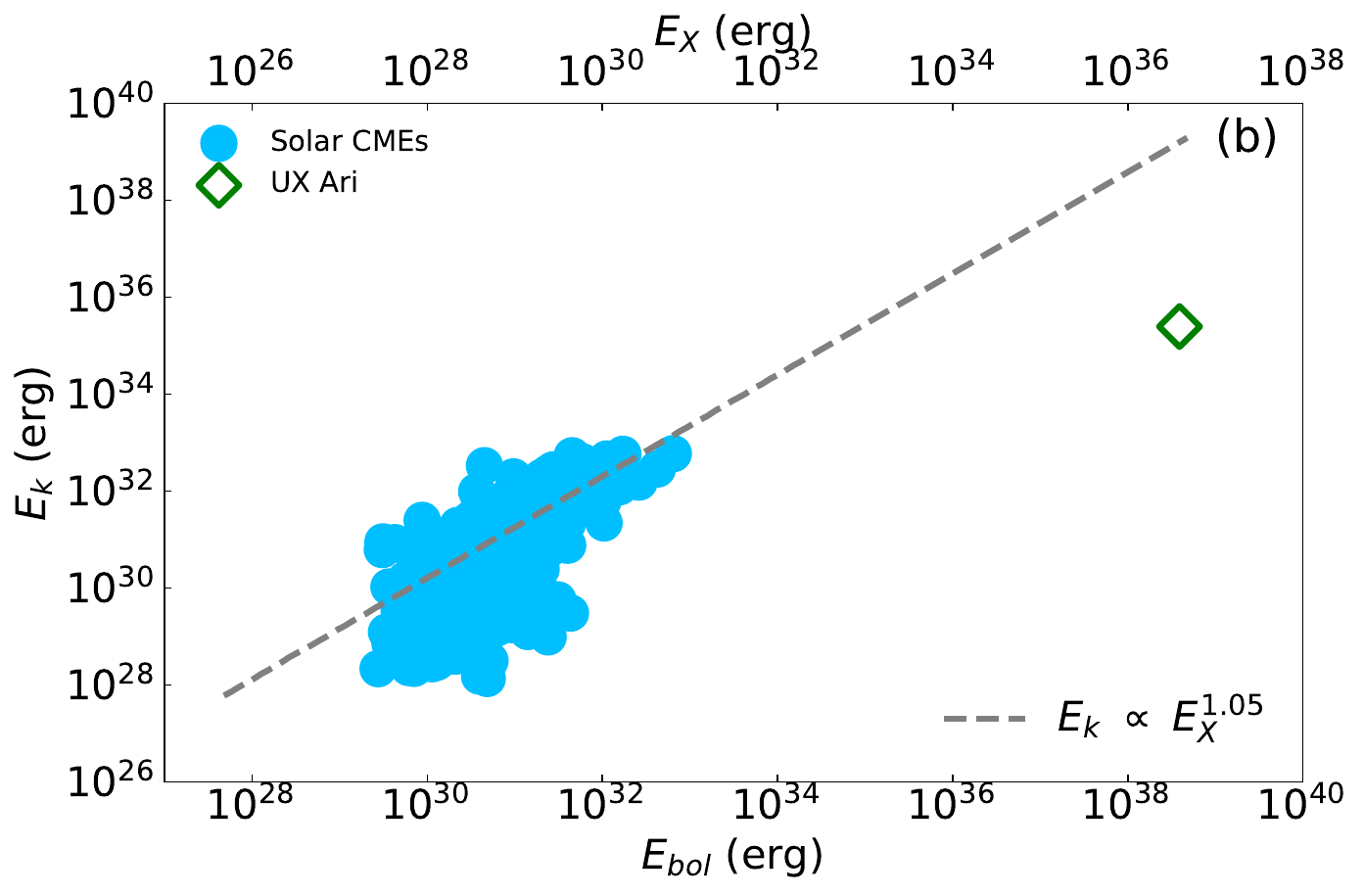}
\caption{Comparison with solar CMEs. (a) CME mass $M_{\mathrm{CME}}$ and (b) kinetic energy $E_{k}$ are plotted as functions of bolometric flare energy (lower horizontal axis) and GOES-band X‑ray energy (upper axis). Solar CME data from \citet{Yashiro2009} are shown as sky‑blue filled circles. The dashed lines represent power‑law fits to the solar CME relations: $M_{\mathrm{CME}} \propto E_{X}^{0.59}$ and $E_{k} \propto E_{X}^{1.05}$ \citep{Drake2013}; the dash-dotted line corresponds to $M_{\mathrm{CME}} \propto E_{X}^{0.7}$ \citep{Aarnio2012}.}
\label{Fig6}
\end{figure}

Figure~\ref{Fig6} compares the detected stellar CME with solar CMEs by plotting its mass and kinetic energy against flare energy in both the GOES 1–8~$\AA$ X-ray band and bolometric emission. The conversion from the H$\alpha$-derived flare energy to the GOES band and subsequently to bolometric energy follows the approach described in \citet{Cao2024}. As shown in the figure, the CME mass generally follows the solar flare--CME relation, whereas its kinetic energy---computed with the deprojected true eruption velocity---lies below the trend extrapolated from solar CMEs. Therefore, the lower kinetic energy cannot be attributed to the projection effect in the eruption velocity; instead, it likely results from prominence eruptions typically having lower velocities than solar CMEs \citep{Maehara2021, Namekata2022}. In addition, the stronger overlying magnetic field on active stars can suppress CME acceleration, leading to reduced speeds and consequently smaller kinetic energies \citep[e.g.,][]{Drake2016, Alvarado2018}.
%------------------------------------------------------------------------------------------%
%------------------------------------------------------------------------------------------%

The long‑term effect of CME on stellar evolution depends on its occurrence rate and mass loss budget. CMEs and stellar winds are the two main channels for mass and angular momentum loss from active stars. The CME mass loss rate can be written as \(\dot{M}_{\mathrm{CME}} = M_{\mathrm{CME}} / \tau\), where \(\tau\) is the average time between CMEs. For UX~Ari, this possible CME has \(M_{\mathrm{CME}} \geq 2.7\times10^{20}\)~g (\(\approx 1.4\times10^{-13} M_\odot\)). The recurrence timescale is not well known, if assuming a recurrence time of about one year as a conservative lower limit (i.e., at least one such event per year on average), this gives \(\dot{M}_{\mathrm{CME}} \gtrsim 1.4\times10^{-13} M_\odot\)~yr\(^{-1}\). This value is comparable to the solar wind mass loss rate (\(\sim 2\times10^{-14} M_\odot\)~yr\(^{-1}\); \citealp{Cohen2011}) within a factor of \(\sim7\), and roughly an order of magnitude higher than the solar CME mass loss rate at solar maximum (\(\sim 10^{-14} M_\odot\)~yr\(^{-1}\); \citealp{Cranmer2017}). Given UX~Ari's much stronger activity, its true CME recurrence rate is likely higher, implying a larger \(\dot{M}_{\mathrm{CME}}\). Direct measurements of UX~Ari's stellar wind are not yet available, but based on its strong coronal and chromospheric emission, the wind is expected to be much stronger than the solar wind. In summary, even our conservative estimate suggests that CME mass loss plays an important role in the evolution of UX~Ari.

The kinetic energy flux at the source region offers a complementary, order‑of‑magnitude measure of the event's intensity. Taking the deprojected kinetic energy of \(E_k = 2.5\times10^{35}\)~erg, a CME source-region area of \(1.9\times10^{22}\)--\(1.9\times10^{23}\)~cm\(^2\) (1\%--10\% of the stellar surface), and a CME eruption duration of \(\sim 1800\)--\(7500\)~s (where the lower bound corresponds to the total exposure time of the first two spectra in which the blueshifted emission is clearly detected, and the upper bound covers the full interval of blueshifted detections), the kinetic energy flux is \(F_{\mathrm{KE}} \approx (0.2\text{--}7)\times 10^{9}\)~erg\,cm\(^{-2}\)\,s\(^{-1}\). Given the substantial uncertainties in the source-region area and duration, this range should be regarded as an order‑of‑magnitude estimate. It is broadly comparable to the typical solar CME source‑region flux of \(\sim 1.7\times 10^{10}\)~erg\,cm\(^{-2}\)\,s\(^{-1}\). Thus, although the total kinetic energy of this stellar CME exceeds that of the largest solar CMEs by about two orders of magnitude, the energy flux per unit area is not dramatically larger. For a hypothetical close‑in planet (e.g., at \(\sim0.05\)~AU), however, the incident energy flux would be hundreds of times greater than that experienced by the Earth during an extreme solar CME, due to the much smaller orbital distance. Therefore, such a CME would significantly enhance atmospheric erosion and could affect the long‑term habitability of a close‑in planet.

\section{Summary and conclusions}\label{sec5}
Based on high-resolution spectroscopic observations of the highly active RS CVn-type star UX Ari, we have detected a long-duration optical flare and identified a possible stellar CME associated with it. Throughout the event, the evolving H$\alpha$ line profile exhibited various asymmetries, reflecting complex plasma dynamics during the flare's evolution. Most notably, around the flare peak, a far-blueshifted emission component likely originating from an erupting prominence is detected. The eruption velocity, especially after correcting for projection effects, is found to significantly exceed the escape velocity of the primary component, providing a strong evidence that the prominence eruption successfully got rid of the stellar gravitational potential and likely developed into a stellar CME. The estimated CME mass is consistent with the solar flare--CME scaling relation, while its kinetic energy lies below the trend extrapolated from solar events. This suggests that although some aspects of stellar CMEs may scale with flare energy in a solar-like manner, others may differ due to different coronal conditions or magnetic field configurations. In addition, the estimated CME mass loss rate and kinetic energy flux, although subjecting to large uncertainties, are of the same order as solar values or higher, suggesting that such events may be non‑negligible for both stellar evolution and exoplanetary habitability.

To improve the accuracy of future stellar CME characterizations, higher time-resolution observations are needed to capture the full evolution process of the eruption from acceleration to propagation, and long-term spectroscopic monitoring is essential to better constrain CME recurrence rates and assess their contribution to stellar mass and angular momentum loss.

%% Please use the acknowledgment and contribution environments. This will 
%% be anonomyized when the "anonymous" style option is used. 
\begin{acknowledgments}
This work is supported by the National Natural Science Foundation of China under grant No. 12288102. We acknowledge the support of the staff of the Xinglong 2.16m telescope. This work is partially supported by National Astronomical Observatories, Chinese Academy of Sciences. The present study is also financially supported by the National Natural Science Foundation of China under grant Nos. 10373023, 10773027, and U1531121, the Yunnan Fundamental Research Projects (grant Nos.~202201AT070186 and 202305AS350009), the Yunnan Revitalization Talent Support Program (Young Talent Project), International Centre of Supernovae, Yunnan Key Laboratory (No.~202302AN360001), and the China Manned Space Program with grant No. CMS-CSST-2025-A15.
\end{acknowledgments}

\appendix
\section{High-resolution Spectroscopic Observing Information for UX~Ari}
Table~\ref{tab2} lists the observing log for UX~Ari. The columns provide the observing date, exposure time (Exp. time), heliocentric Julian date (HJD), and the corresponding orbital phase. Orbital phases are computed using the ephemeris HJD = 2,456,238.134 + $6.437888P$ from \citet{Hummel2017}. Phase zero is defined as the inferior conjunction of the G5~V secondary star passing in front of the K0~IV primary star.

\begin{deluxetable}{lccclccc}
\tablenum{A1}
\tablecaption{Observing Log of UX~Ari\label{tab2}}
\tablewidth{0pt}
\setlength{\tabcolsep}{7pt} 
\tablehead{
\colhead{UT date} & \colhead{Exp. time} & \colhead{HJD} & \colhead{Orbital phase} & \colhead{UT date} & \colhead{Exp. time} & \colhead{HJD} & \colhead{Orbital phase}\\
\nocolhead{} & \colhead{(s)} & \colhead{(2,450,000+)} & \nocolhead{} & \nocolhead{} & \colhead{(s)} & \colhead{(2,450,000+)} & \nocolhead{}
}
\startdata
2015 Oct 27 & 900 & 7323.312 & 0.561 & 2015 Oct 30 & 900 & 7326.371 & 0.036\\
2015 Oct 27 & 900 & 7323.325 & 0.563 & 2015 Oct 31 & 900 & 7327.296 & 0.180\\
2015 Oct 27 & 900 & 7323.338 & 0.565 & 2015 Oct 31 & 900 & 7327.309 & 0.182\\
2015 Oct 27 & 900 & 7323.351 & 0.567 & 2015 Oct 31 & 900 & 7327.345 & 0.188\\
2015 Oct 28 & 900 & 7324.343 & 0.721 & 2015 Oct 31 & 900 & 7327.358 & 0.190\\
2015 Oct 28 & 900 & 7324.356 & 0.723 & 2015 Oct 31 & 900 & 7327.370 & 0.192\\
2015 Oct 28 & 900 & 7324.369 & 0.725 & 2015 Oct 31 & 900 & 7327.383 & 0.194\\
2015 Oct 28 & 900 & 7324.382 & 0.727 & 2015 Nov 1  & 900 & 7328.283 & 0.333\\
2015 Oct 29 & 900 & 7325.279 & 0.867 & 2015 Nov 1  & 900 & 7328.296 & 0.335\\
2015 Oct 29 & 900 & 7325.292 & 0.869 & 2015 Nov 1  & 900 & 7328.328 & 0.340\\
2015 Oct 29 & 900 & 7325.305 & 0.871 & 2015 Nov 1  & 900 & 7328.341 & 0.342\\
2015 Oct 29 & 900 & 7325.317 & 0.873 & 2015 Nov 1  & 900 & 7328.373 & 0.347\\
2015 Oct 29 & 900 & 7325.357 & 0.879 & 2015 Nov 1  & 1200& 7328.387 & 0.350\\
2015 Oct 29 & 900 & 7325.369 & 0.881 & 2015 Nov 2  & 900 & 7329.314 & 0.493\\
2015 Oct 29 & 900 & 7325.382 & 0.883 & 2015 Nov 2  & 900 & 7329.327 & 0.495\\
2015 Oct 29 & 900 & 7325.395 & 0.885 & 2015 Nov 2  & 750 & 7329.362 & 0.501\\
2015 Oct 30 & 900 & 7326.332 & 0.030 & 2015 Nov 2  & 900 & 7329.374 & 0.503\\
2015 Oct 30 & 900 & 7326.345 & 0.032 & 2015 Nov 2  & 900 & 7329.387 & 0.505\\
2015 Oct 30 & 900 & 7326.358 & 0.034 & 2015 Nov 2  & 900 & 7329.400 & 0.507\\
\enddata
\tablecomments{HJD and orbital phase are calculated for the mid-exposure of each observation.}
\end{deluxetable}

%% For this sample we use BibTeX plus aasjournalv7.bst to generate the

\bibliography{sample701}{}
\bibliographystyle{aasjournalv7}



\end{document}